\documentclass[showpacs,preprintnumbers,amsmath,amssymb,superscriptaddress,nofootinbib,english]{revtex4}

\usepackage{epstopdf}
\usepackage{graphicx}
\usepackage{dcolumn}
\usepackage{bm}
\usepackage{epsfig}
\usepackage{graphicx}
\usepackage{hyperref}
\usepackage[usenames]{color}
\usepackage{url}

\hypersetup{
    colorlinks=true,
    linkcolor=green,
    citecolor=blue,
}

\newcommand{\remove}[1]{}

\def\be{\begin{equation}}
\def\ee{\end{equation}}
\def\ba{\begin{eqnarray}}
\def\ea{\end{eqnarray}}

\begin{document}

\title{Modified gravity and the radiation dominated epoch}
\author{Carsten van de Bruck}
\email[Email address: ]{C.vandeBruck@sheffield.ac.uk}
\affiliation{Consortium for Fundamental Physics, School of Mathematics and Statistics, University of Sheffield, Hounsfield Road, Sheffield, S3 7RH, United Kingdom}

\author{Gregory I. Sculthorpe}
\email[Email address: ]{app09gis@sheffield.ac.uk}
\affiliation{Consortium for Fundamental Physics, School of Mathematics and Statistics, University of Sheffield, Hounsfield Road, Sheffield, S3 7RH, United Kingdom}

\date{\today}

\begin{abstract}
In this paper we consider scalar-tensor theories, allowing for both conformal and disformal couplings to a fluid with a generic equation of state.
We derive the effective coupling for both background cosmology and for perturbations in that fluid. As an application we consider the scalar degree of freedom to be coupled to baryons and study the dynamics of the tightly coupled photon-baryon fluid in the early 
universe. We derive an expression for the effective speed of sound, which differs from its value in General Relativity. We apply our findings to the $\mu$--distortion of 
the cosmic microwave background radiation, which depends on the effective sound-speed of the photon-baryon fluid, and show that the predictions differ 
from General Relativity. Thus, the $\mu$--distortion provides further information about gravity in the very early universe well before decoupling. 

\end{abstract}

\maketitle

\section{Introduction}
In the search for an explanation of the observed accelerated expansion of the Universe, theorists have questioned the validity of General Relativity. While the cosmological constant is the easiest explanation and in very good agreement with observations so far, its magnitude and origin remain a mystery and other explanations are searched for. Theories such as $f(R)$-theories, massive gravity, galileon models and other interacting dark energy models have been studied in much detail and their phenomenological consequences explored. Most of these theories give rise to deviations from the $\Lambda$CDM model which can be looked for with future observations and therefore models will be discarded or heavily constrained. We refer to \cite{Amendola} and \cite{Copeland} for excellent reviews. 

In this paper we consider scalar-tensor theories of gravity as an extension of General Relativity. These theories have been revived recently, either because they are equivalent to some theories mentioned above (for example $f(R)$ models can be written in the form of a scalar-tensor theory) or they describe limits of those theories. Here, we consider not only conformal couplings of matter to the scalar field but also disformal couplings, as motivated for example from galileon models, certain limits of massive gravity and Lorentz-breaking models of gravity \cite{Brax0,KoivistoMotaZumalacarregui12}. As we will see, the effective couplings of matter to the scalar degree of freedom are much more complicated expressions and depend not only on the scalar field itself but also on its time variation. Such models motivate time-varying effective couplings.  

Unfortunately, it has recently been shown that disformal couplings are difficult to constrain with local experiments (see e.g. \cite{Brax0,Noller,Brax1}).
Therefore, as an application, we study the consequences of modifications of gravity in the radiation dominated epoch. In particular, we find a generic expression for the sound-speed of the tightly coupled photon-baryon fluid in theories with conformal and disformal couplings and calculate the distortion of the cosmic microwave background (CMB) caused by the dissipation of acoustic waves. As is well known, the dissipation of acoustic waves injects energy into the photons and therefore produces slight deviations from the blackbody spectrum by producing a positive chemical potential $\mu$. The deviations are small; the chemical potential created by the dissipation of acoustic waves is of order $\mu \approx 10^{-8}$ in the standard inflationary scenario and is not in violation of the current constraint $|\mu|<9\times 10^{-5}$ (for work on these issues, see e.g.\cite{HuSilk93,HuScottSilk94,KhatriSunyaevChluba12,KhatriSunyaevChluba12b,PajerZaldarriaga12} and references therein). Proposed experiments such as PIXIE, however, reach this sensitivity to search for deviations of the order $\mu\approx 10^{-8}$ \cite{pixie}. As such, these observations probe the primordial power spectrum at very small scales and constrain the inflationary epoch \cite{Easson,Chlubainflation}. We point out that in general modifications of gravity change the spectral distortions because of the different sound-speed of the coupled photon-baryon plasma. For theories in which the field is very heavy the sound-speed is smaller than in General Relativity and thus an absence of a $\mu$-type distortion in the CMB spectrum could be explained by a lower sound-speed during the epoch when the distortion is created ($5\times 10^4 < z < 2\times 10^6$). However, the sound horizon is well constrained by measurements of the CMB anisotropies (e.g. the position of the first peak is well known), and, as we will see, this further constrains the type of modified gravity theories which are allowed. It is worth pointing out that the $\mu$--distortion is the earliest direct probe of modifications of gravity. 

The paper is organised as follows: in Section 2 we present the action considered in this paper, write down the perturbation equations which govern the dynamics of the coupled photon-baryon fluid and derive the effective coupling between the scalar field and matter and the effective sound-speed for the coupled photon-baryon fluid. In Section 3 we calculate the $\mu-$distortion. In Section 4 we present our conclusions and an outlook of future work.

\section{Evolution of Perturbations}

\subsection{General Equations}

The action we are considering is of scalar-tensor form, namely
\be
{\cal S} = \int \sqrt{-g}d^4 x\left[ \frac{{\cal R}}{16 \pi G} -\frac{1}{2}g^{\mu\nu}(\partial_\mu\phi)(\partial_\nu\phi) - V(\phi) \right] + S_{\rm matter} (\chi_i,{\tilde g}_{\mu\nu}^{(i)}),
\ee
in which ${\cal R}$ is the Ricci scalar, $\chi_i$ are the matter fields in the theory (relativistic and non-relativistic), $\phi$ is an additional scalar degree of freedom and the metrics $\tilde{g}^{(i)}$ are related to the metric $g$ by
\be\label{disformalmetric}
{\tilde g}_{\mu\nu}^{(i)} = C^{(i)}(\phi)g_{\mu\nu}+D^{(i)}(\phi)\partial_\mu \phi \partial_\nu \phi~.
\ee
Each matter species can, in general, couple to a different metric, in which the couplings are described by the functions $C(\phi)$ (called the conformal factor) and $D(\phi)$ (called the disformal factor). However, multiple couplings greatly complicate the equations of motion and in this paper we will consider a coupling to a single species only. We will return to multiple couplings in future work. 

The field equations can easily be obtained from the action above \cite{KoivistoMotaZumalacarregui12}. The scalar field equation is given by
\be
g^{\mu\nu} \nabla_\mu \nabla_\nu \phi  - \frac{d V}{d\phi} + Q= 0,
\ee
with 
\be
Q = \frac{C'}{2C}T^\mu_{~\mu} -\nabla_\nu \left(\frac{D}{C}\phi_{,\mu}T^{\mu\nu}\right) +\frac{D'}{2C}\phi_{,\mu}\phi_{,\nu}T^{\mu\nu}
\ee
and $T^{\mu\nu}$ being the energy-momentum tensor of the species coupled to $\phi$, which is consequently not conserved:
\be\label{emconservation}
\nabla_\mu T^{\mu}_{\phantom{1}\nu} = Q\phi_{,\nu} .
\ee
We will write down the equations for the perturbations only and work in the conformal (Newtonian) gauge in which the metric is given by
\be
ds^2 = a^2(\eta)[-(1+2\Psi)d\eta^2 + (1-2\Phi)\delta_{ij}dx^i dx^j ].
\ee
Perturbing equation (\ref{emconservation}) yields the following equations
\begin{align}
\dot{\delta}_i = \, & -(1+w_{i})(\theta_i-3\dot{\Phi}) -3{\cal H}\left(\frac{\delta P_{i}}{\delta\rho_{i}}-w_{i}\right)\delta_{i}+\frac{Q_0}{\rho_i}\dot{\phi}\,\delta_{i}-\frac{Q_0}{\rho_i}\delta\dot{\phi}-\frac{\dot{\phi}}{\rho_i}\delta Q, \label{one}\\
\dot{\theta}_i = \, & -{\cal H}(1-3w_{i})\theta_{i}-\frac{\dot{w_{i}}}{1+w_{i}}\theta_{i}+k^2\Psi+\frac{\delta P_{i}/\delta\rho_{i}}{1+w_{i}}k^{2}\delta_{i}-k^{2}\sigma_{i}+\frac{Q_0}{\rho_i}\dot{\phi}\theta_{i}-\frac{Q_0}{(1+w_{i})\rho_i}k^2\delta\phi, \label{two}
\end{align}
whilst the scalar field perturbations obey the Klein-Gordon equation
\be
\delta\ddot{\phi}+2{\cal H}\delta\dot{\phi}+(k^{2}+a^{2}V'')\delta\phi = \dot{\phi}(\dot{\Psi}+3\dot{\Phi})-2a^{2}(V'-Q_{0})\Psi+a^{2}\delta Q.
\ee
The zero-order part of $Q$ is
\be\label{Qbackgroundcoupling}
Q_0 = -\frac{a^{2}C'(1-3w_{i})-2D(3{\cal H}\dot{\phi}(1+w_{i})+a^{2}V'+\frac{C'}{C}\dot{\phi}^{2})+D'\dot{\phi}^{2}}{2(a^{2}C+D(a^2\rho_{i}-\dot{\phi}^{2}))}\,\rho_{i} \,,
\ee
and the perturbation of $Q$ is
\be\label{Qperturb}
\delta Q = -\frac{\rho_{i}}{a^{2}C+D(a^{2}\rho_{i}-\dot{\phi}^2)}[{\cal B}_1\delta_{i}+{\cal B}_2\dot{\Phi}+{\cal B}_3\Psi+{\cal B}_4\delta\dot{\phi}+{\cal B}_5\delta\phi],
\ee
where
\begin{align}
{\cal B}_1 = \, & \frac{a^{2}C'}{2}\left(1-3\frac{\delta P_{i}}{\delta\rho_{i}}\right)-3D{\cal H}\dot{\phi}\left(1+\frac{\delta P_{i}}{\delta\rho_{i}}\right)-Da^{2}(V'-Q_{0})-D\dot{\phi}^2\left(\frac{C'}{C}-\frac{D'}{2D}\right), \\
{\cal B}_2 = \, & 3D\dot{\phi}(1+w_{i}), \\
{\cal B}_3 = \, & 6D{\cal H}\dot{\phi}(1+w_{i})+2D\dot{\phi}^2\left(\frac{C'}{C}-\frac{D'}{2D}+\frac{Q_0}{\rho_{i}}\right), \\
{\cal B}_4 = \, & -3D{\cal H}\dot{\phi}(1+w_{i})-2D\dot{\phi}\left(\frac{C'}{C}-\frac{D'}{2D}+\frac{Q_0}{\rho_{i}}\right), \\
{\cal B}_5 = \, & \frac{a^{2}C''(1-3w_{i})}{2}-Dk^{2}(1+w_{i})-Da^{2}V''-D'a^{2}V'-3D'{\cal H}\dot{\phi}(1+w_{i}) \nonumber \\ & -D\dot{\phi}^2\left(\frac{C''}{C}-\left(\frac{C'}{C}\right)^{2}+\frac{C'D'}{CD}-\frac{D''}{2D}\right)+(a^{2}C'+D'a^2\rho_{i}-D'\dot{\phi}^{2})\frac{Q_0}{\rho_{i}},
\end{align}
and the subscript $i$ denotes the species the field is coupled to. The expression for $Q_0$ agrees with \cite{KoivistoMotaZumalacarregui12} for the case $w_i=0$.

For the rest of this paper we shall look at the case when the scalar field is coupled to baryons only. We are treating photons and baryons as fluids, coupled via Thomson scattering. For photons and baryons, Eqns. (\ref{one}) and (\ref{two}) become 

\begin{align}
\dot{\delta}_\gamma = \, & -\tfrac{4}{3}\theta_\gamma +4\dot{\Phi},  \\
\dot{\theta}_\gamma = \, & \tfrac{1}{4}k^2\delta_\gamma -k^2\sigma_\gamma +k^2\Psi+an_e\sigma_{T}(\theta_{b}-\theta_{\gamma}), \\
\dot{\delta}_b = \, & -\theta_b +3\dot{\Phi}+\frac{Q_0}{\rho_b}\dot{\phi}\,\delta_{b}-\frac{Q_0}{\rho_b}\delta\dot{\phi}-\frac{\dot{\phi}}{\rho_b}\delta Q, \\
\dot{\theta}_b = \, & -{\cal H}\theta_{b}+k^2\Psi+\frac{an_e\sigma_T}{R}(\theta_{\gamma}-\theta_{b})+\frac{Q_0}{\rho_b}\dot{\phi}\theta_{b}-\frac{Q_0}{\rho_b}k^2\delta\phi ~,
\end{align}
where we have added the interaction terms for Thomson scattering and $R=3\rho_b/4\rho_\gamma$.

\subsection{Tight-Coupling Approximation}
 
We are interested in scales much smaller than the horizon and on time-scales much smaller than the Hubble expansion rate. To derive a second order differential equation for $\delta_\gamma$, we ignore therefore terms which involve the Hubble expansion rate, the time-evolution of the background scalar field, and the time-derivatives of the scalar field perturbations and the gravitational potential. In this limit, the relevant equations read
\begin{align}
\dot{\delta}_\gamma = \, & -\tfrac{4}{3}\theta_\gamma, \label{sHdeltagammadot} \\
\dot{\theta}_\gamma = \, & \tfrac{1}{4}k^2\delta_\gamma -k^2\sigma_\gamma +k^2\Psi+an_e\sigma_{T}(\theta_{b}-\theta_{\gamma}), \label{sHthetagammadot} \\
\dot{\delta}_b = \, & -\theta_b, \label{sHdeltabdot} \\
\dot{\theta}_b = \, & k^2\Psi+\frac{an_e\sigma_T}{R}(\theta_{\gamma}-\theta_{b})-\frac{Q_0}{\rho_b}k^2\delta\phi, \label{sHthetabdot} \\
(k^2 + a^2 V'')\delta\phi = \, & a^{2}\delta Q, \label{sHKG}
\end{align}
where
\be
Q_0 = \frac{2DV'-C'}{2(C+D\rho_{b})}\rho_{b},\qquad \delta Q = -\frac{\rho_{b}}{a^{2}(C+D\rho_{b})}[{\cal B}_1\delta_{b}+{\cal B}_5\delta\phi],
\ee
and
\begin{align}
{\cal B}_1 = \, & \frac{a^{2}C'}{2}-Da^{2}(V'-Q_{0}), \\
{\cal B}_5 = \, & \frac{a^{2}C''}{2}-Dk^{2}-Da^{2}V''-D'a^{2}V'+a^{2}(C'+D'\rho_{b})\frac{Q_0}{\rho_{b}}.
\end{align}
From equation (\ref{sHKG}) we find
\be
\delta\phi = -\frac{{\cal B}_1\rho_b}{(C+D\rho_{b})(k^2 +a^2 V'')+{\cal B}_5\rho_{b}}\delta_{b}.\label{deltaphi}
\ee
To leading order in the tight-coupling approximation $\dot{\tau}\equiv an_{e}\sigma_{T}\to\infty$ which implies
\be
\theta_\gamma \approx \theta_b ,\label{tightcouplapprox}
\ee
and therefore
\be
\delta_b \approx \tfrac{3}{4}\delta_\gamma .\label{adiabaticity}
\ee
Additionally it can be shown (e.g. \cite{PeterUzan}) that
\be
\sigma_{\gamma} = \frac{16}{45\dot{\tau}}\theta_{\gamma},
\ee
and so we can neglect the anisotropic stress. Using equation (\ref{tightcouplapprox}) on the left-hand side of equation (\ref{sHthetabdot}) we have
\be
\dot{\tau}(\theta_\gamma -\theta_b) = R\left(\dot{\theta}_\gamma -k^{2}\Psi+\frac{Q_0}{\rho_b}k^2\delta\phi\right),
\ee
which can be inserted in equation (\ref{sHthetagammadot}) to get
\be
\dot{\theta}_\gamma = \tfrac{1}{4}k^2\delta_\gamma +k^2\Psi+R\left(\dot{\theta}_\gamma -k^{2}\Psi+\frac{Q_0}{\rho_b}k^2\delta\phi\right).
\ee
Rearranging and using equation (\ref{sHdeltagammadot}) we find
\be
\ddot{\delta}_\gamma = -\frac{k^{2}}{3(1+R)}\delta_{\gamma}-\frac{4}{3}k^{2}\Psi+\frac{Q_0}{\rho_b}\frac{4Rk^2}{3(1+R)}\delta\phi.
\ee
We can then substitute for $\delta\phi$ using equations (\ref{deltaphi}) and (\ref{adiabaticity}) and this leads to
\be
\ddot{\delta}_\gamma + (1+3R{\cal F})k^{2}c_{s}^{2}\delta_\gamma = -\frac{4}{3}\Psi,
\ee
where $c_{s}=1/\sqrt{3(1+R)}$ is the standard sound-speed and
\be
{\cal F} = \frac{Q_{0}{\cal B}_1}{(C+D\rho_{b})(k^2 +a^2 V'')+{\cal B}_5\rho_{b}}.
\ee
From this equation we can read off the modified sound-speed
\be
{\tilde c}_{s}^2 = c_{s}^2(1+3R{\cal F}),
\ee
which reduces to the expression given in \cite{Brax2} for the purely conformal case.

To account for Silk damping it is necessary to go beyond leading order in the tight-coupling approximation. Doing so (see the Appendix for a derivation) yields
\be
\delta_\gamma \propto e^{ik{\tilde r}_{s}}e^{-k^{2}/{\tilde k}_{D}^{2}}
\ee
where
\be
{\tilde r}_{s} = \int_{0}^{\eta}{\tilde c}_{s}d\eta'
\ee
is the sound horizon (which in general differs from its value in General Relativity) and ${\tilde k}_D$ is the modified damping wavenumber:
\be
\frac{1}{{\tilde k}_D^2} = \int_z^{\infty}\frac{dz(1+z)}{6H(1+R)n_{e}\sigma_T}\left[\frac{16}{15}+\frac{R^{2}}{1+R}\left(1-3(2+R){\cal F}+\frac{3(1+R)}{{\tilde c}_{s}^{2}}{\cal F}^{2}\right)\right].
\ee
It can be seen that the additional terms are multiplied by $R^{2}/(1+R)$. Deep in the radiation dominated epoch when $\rho_\gamma \gg \rho_b$ these terms will be totally insignificant. We found that these modifications are irrelevant for the $\mu$--distortion considered in the next section.

The evolution of ${\tilde c}_s$ depends on the details of the coupling functions $C(\phi)$ and $D(\phi)$ and can be very complicated, even in purely conformal theories $(D=0)$. Potentially, ${\tilde c}_s^2$ can become negative, signalling instabilities. We will, in this paper, not study the consequences of an imaginary sound-speed during the radiation dominated epoch, although it would be interesting to study black hole formation on small scales in these scenarios. 

There are several observables which depend on the sound-speed and can therefore be used to search for deviations from General Relativity. In this paper we consider two of them, namely the sound horizon at decoupling and the $\mu$--distortion. The sound horizon at decoupling determines, for example, the position of the peaks in the anisotropy spectrum. Since the position of the first peak is well known, ${\tilde r}_s(z_{\rm dec})$ can not vary too much from its value in General Relativity. As an integral of ${\tilde c}_s$ over time, ${\tilde r}_s(z_{\rm dec})$ is dependent mostly on the evolution of ${\tilde c}_s$ for redshifts below $10^5$ or so. The $\mu$--distortion of the CMB blackbody spectrum is created in the redshift range $5\times 10^4 \leq z \leq 2\times 10^6$ and probes length scales of order $k \approx 10$ Mpc$^{-1}$ to $k \approx 10^4$ Mpc$^{-1}$ . As such, $\mu$ does not only provide useful information about the primordial curvature perturbation and therefore about about inflationary physics, but also about modifications of gravity. In fact, the $\mu$--distortion of the CMB spectrum is the earliest possible direct probe of modifications of gravity available to us. We therefore turn our attention to calculate $\mu$ in the next section. 

\section{$\mu$-type distortion due to dissipation of acoustic waves}

The evolution of the $\mu$--distortion is given by \cite{HuSilk93,HuScottSilk94}
\be
\frac{d\mu}{dt} = -\frac{\mu}{t_{\rm DC}(z)}+1.4\frac{d{\cal Q}/dt}{\rho_{\gamma}},
\ee
where the last term describes the change of $\mu$ due to the input of energy into the coupled photon-baryon fluid and the first term describes the thermalization process with $t_{\rm DC}$ being the double Compton scattering time scale. We will give the expression for ${\cal Q}$ below. 

The solution of this equation is 
\begin{align}
\mu = \, & 1.4\int_{t(z_{1})}^{t(z_{2})}dt\frac{d{\cal Q}/dt}{\rho_{\gamma}}e^{-(z/z_{\rm DC})^{5/2}} \nonumber \\ = \, & 1.4\int_{z_{1}}^{z_{2}}dz\frac{d{\cal Q}/dz}{\rho_{\gamma}}e^{-(z/z_{\rm DC})^{5/2}}, \label{mu}
\end{align}
where
\be
z_{\rm DC} = 1.97\times10^{6}\left(1-\frac{1}{2}\left(\frac{Y_{p}}{0.24}\right)\right)^{-2/5}\left(\frac{\Omega_{b}h^{2}}{0.0224}\right)^{-2/5},
\ee
and $Y_{p}$ is the primordial helium mass fraction.

To calculate the energy input, we follow \cite{KhatriSunyaevChluba12b,PajerZaldarriaga12} who showed that the energy density of an acoustic wave in the photon-baryon plasma can be written as 
\be\label{energydensity}
{\cal Q} = \rho_\gamma \frac{c_s^2}{1+w_{\gamma}} \left<  \delta_\gamma^2(\vec x) \right>_P,
\ee
where we here ignore the baryon density (which is much smaller than the energy density in photons $\rho_\gamma$), 
$c_s$ is the sound-speed of the wave, $\delta_\gamma$ is the photon density contrast, $w_{\gamma}=p_{\gamma}/\rho_{\gamma}=\tfrac{1}{3}$ and the average $\left< ...  \right>_P$ denotes an average over one period of oscillation. We have 
\be
\left< \delta_\gamma^2(\vec x) \right>_P = \int \frac{d^{3}k}{(2\pi)^3}P_\gamma(k),
\ee
where $P_\gamma (k)$ is the power spectrum of the photon density contrast. The power spectrum $P_\gamma(k)$ for scales 
well within the horizon can be related to the primordial power spectrum $P_\gamma^{i}(k)$ by
\be
P_\gamma(k) = \Delta_{\gamma}^{2}(k)P_\gamma^{i}(k),
\ee
where $\Delta_\gamma$ is the transfer function. In General Relativity, the transfer function reads 
\be
\Delta_{\gamma}(k) = 3{\rm cos}(k{\tilde r}_{s})e^{-k^{2}/k_{D}^{2}},
\ee
where ${\tilde r}_{s}=\int {\tilde c}_s d\eta $ is the sound horizon and $k_{D}$ is the diffusion scale. The factor of 3 takes into account for the fact that the potential is decaying on super-horizon scales, leading to an enhancement in $\delta_\gamma$ \cite{HuSugiyama,KhatriSunyaevChluba12b}. It is here that modifications of gravity could play an important role too: if the field is light ($m<H$) the enhancement might be larger or smaller, depending on the coupling functions $C$ and $D$. On the other hand, in theories such as chameleon theories, the mass of the scalar degree of freedom is much larger than the Hubble expansion rate $H$ and the field is short ranged. This is the case we are considering in this paper and so we expect the factor of 3 to be a very good approximation in the theories we consider here. We will return to the case $m<H$ in a separate publication \cite{usfuture} in which we discuss the transfer functions for other types of theories of modified gravity.

The primordial power spectrum can be written \cite{KhatriSunyaevChluba12}
\be
P_\gamma^{i}(k) \approx 1.45P_\zeta = 1.45A_{\zeta}\frac{2\pi^{2}}{k^{3}}\left(\frac{k}{k_{0}}\right)^{n_{s}-1+\tfrac{1}{2}\alpha {\rm ln}\left(\tfrac{k}{k_{0}}\right)}
\ee
where $A_{\zeta}={\rm 2.4\times10^{-9}}$, $k_{0}={\rm 0.002Mpc^{-1}}$, $n_s$ is the spectral index and $\alpha \equiv dn_{s}/d{\rm ln}k$ is the running of the index.

In the standard case ${\tilde c}_s$ is independent of the wavenumber $k$, but as we have seen, the interaction of the baryons with the scalar field causes ${\tilde c}_s$ to be dependent on the wavenumber. We are interested in scales much smaller than the interaction range of the scalar field $(k\gg m(a)a)$ and ${\tilde c}_s$ is only varying very slowly in time. Using equation (\ref{energydensity}) for a wave with a given wave vector $\vec k$, one can derive the energy density of that wave. Integrating then over all waves gives the total energy density 
\be
{\cal Q} = \tfrac{3}{4}\rho_\gamma \int \frac{d^3 k}{(2\pi)^3} {\tilde c}^2_s(k) \Delta_{\gamma}^{2}(k)P_\gamma^{i}(k)~.
\ee

Then using the fact that the average over one oscillation of $\cos^2(x)$ is $1/2$ and treating the photon density as effectively constant over the timescales considered we can obtain the energy release per unit redshift 
\be
\frac{d{\cal Q}/dz}{\rho_\gamma} =  1.1745\times10^{-8} \int \frac{dk}{k}\left(\frac{k}{k_{0}}\right)^{n_{s}-1+\tfrac{1}{2}\alpha_{s}{\rm ln}\left(\tfrac{k}{k_{0}}\right)}\frac{d}{dz}\left({\tilde c}^2_s(k) e^{-2k^{2}/k_{D}^{2}}\right). 
\ee
We can then use this expression to evaluate equation (\ref{mu}). 

For the models considered here the effect of modified gravity on $\mu$ comes solely from the modified sound-speed. How $\tilde c_s$ deviates from its evolution in General Relativity depends on the coupling functions $C(\phi)$ and $D(\phi)$ and the potential $V(\phi)$. Therefore, a plethora of possibilities could be explored. To be specific however, in this paper we focus on the purely conformal case for which (with $C=e^{2\beta\phi}$ and $D=0$) the effective sound-speed can be written as 
\be
\tilde c_s^2= c_s^2  \left(1-\frac{9\Omega_b \beta^2 R{\cal H}^2}{k^2 +m^2a^2}\right).
\ee
In this case the effective sound-speed is {\it smaller} than in General Relativity. Precisely when and how much smaller is dictated by how the coupling stength $\beta$ and the mass ($m^{2}=d^{2}V/d\phi^{2}$) evolve in time. It is instructive to estimate the deviation of $\tilde c_s^2$ from its value in General Relativity. We find 
\be\label{devA}
{\cal A} \equiv \frac{9\Omega_b \beta^2 R{\cal H}^2}{k^2 +m^2a^2}  = \frac{27\Omega_{b,0}^2 H_0^2\beta^2}{4\Omega_{\gamma,0}(k^2 + a^2 m^2)}
\ee
Looking at a regime for which $a^2 m^2 \ll k^2$ and $k = 100$ Mpc$^{-1}$, we find ${\cal A}=10^{-3}$ for $\beta=10^3$, and ${\cal A}=0.15$ for 
 $\beta=10^4$. Therefore, rather large couplings are needed in this case for the sound-speed to deviate significantly from its value in General Relativity. 
As mentioned above, for such a theory to be consistent with the observed CMB anisotropies, the sound horizon cannot be modified very much. Therefore ${\cal A}$ has to become small before decoupling which requires that either $\beta$ decreases or $m$ increases well before decoupling. In the example that follows we focus on the former case for which $\beta$ becomes smaller in time and as a concrete example we consider 

\be
\beta = b\left[1 + \tanh\left(  d(z - z_0)   \right)\right],
\ee
where $z_0$ is a redshift after which the coupling becomes rapidly smaller, and $b$ and $d$ are constants. For the mass we assume that 

\be
m(z) = m_{\rm rat} H(z),
\ee
with $m_{\rm rat}$ being a constant. In Fig. 1 we show the evolution of ${\tilde c}_s$ for a couple of choices for $b, d, z_0$ and $m_{\rm rat}$. In the first case, the sound-speed deviates from its value in General Relativity at high redshifts ($z>10^6$), but approaches the standard value quickly for $z<10^6$. In the second case, the deviation is quite large for $z>400000$, so ${\tilde c}_s$ approaches $c_s$ at a later time than in the first example and the maximum deviation is greater in this case. In both cases the sound-speed approaches its standard value in General Relativity. A motivation for the behaviour of $\beta$ could be that General Relativity is an attractor in the radiation dominated epoch (see e.g. \cite{Damour} for early ideas on such models). 

\begin{figure}
\begin{center}
\includegraphics[width=8.75cm]{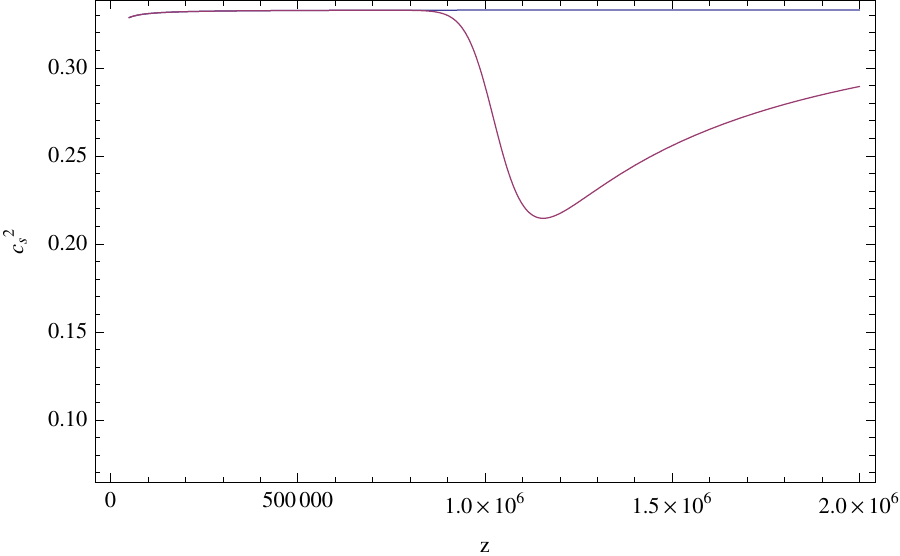}~~~~~~~\includegraphics[width=8.75cm]{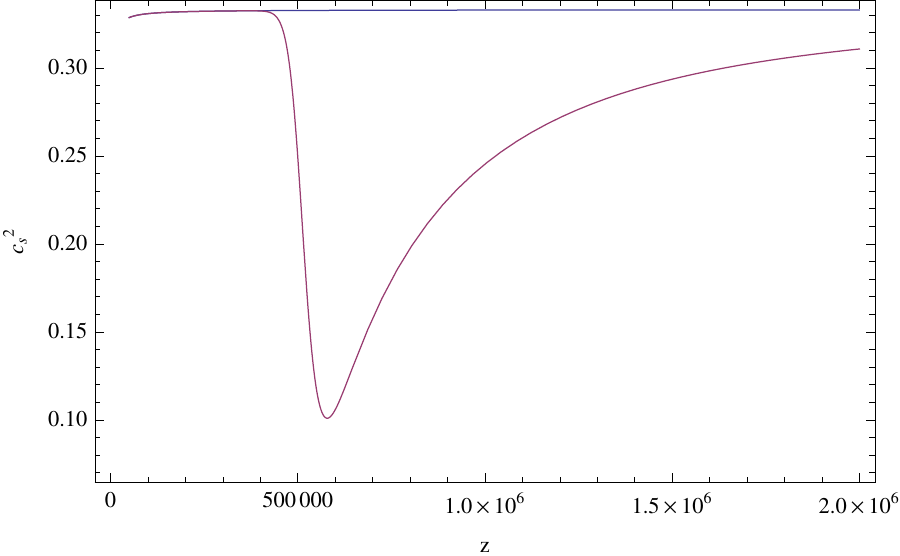}
\end{center}
\caption{These graphs show the evolution of the effective sound-speeds, with $k=100$ Mpc$^{-1}$, for two examples. In the left plot we have taken $b=7\times10^4$, $d = 10^{-5}$, $z_0 = 10^6$ and $m_{\rm rat} =350$. In the right plot we have taken $b=5\times 10^4$, $d = 2\times 10^{-5}$, $z_0 = 5\times 10^5$ and $m_{\rm rat} =350$. We also plot the evolution of the standard sound-speed (upper curve in both graphs).}
\end{figure}

\begin{figure}
\begin{center}
\includegraphics[width=7.5cm]{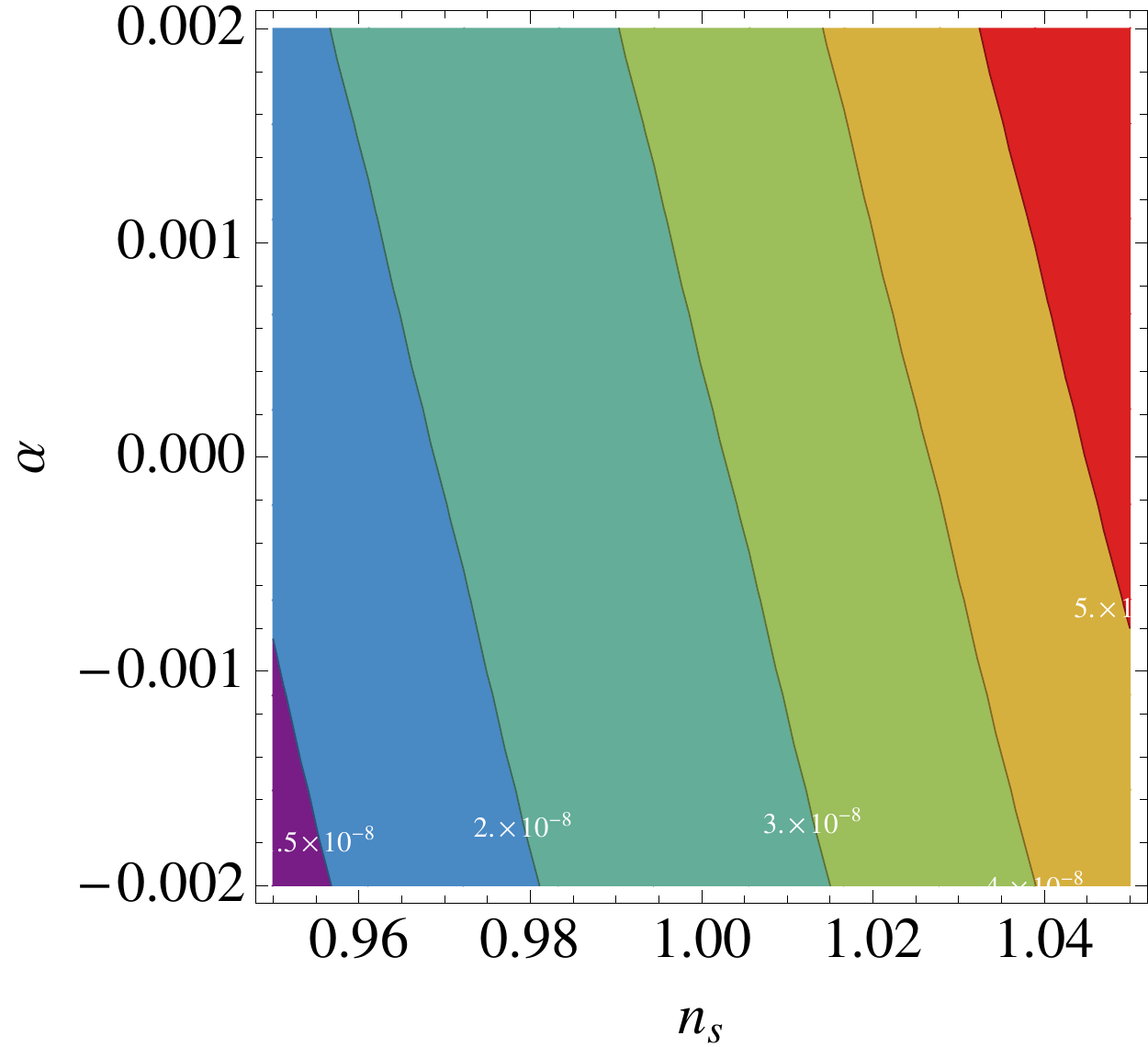}

\includegraphics[width=8cm]{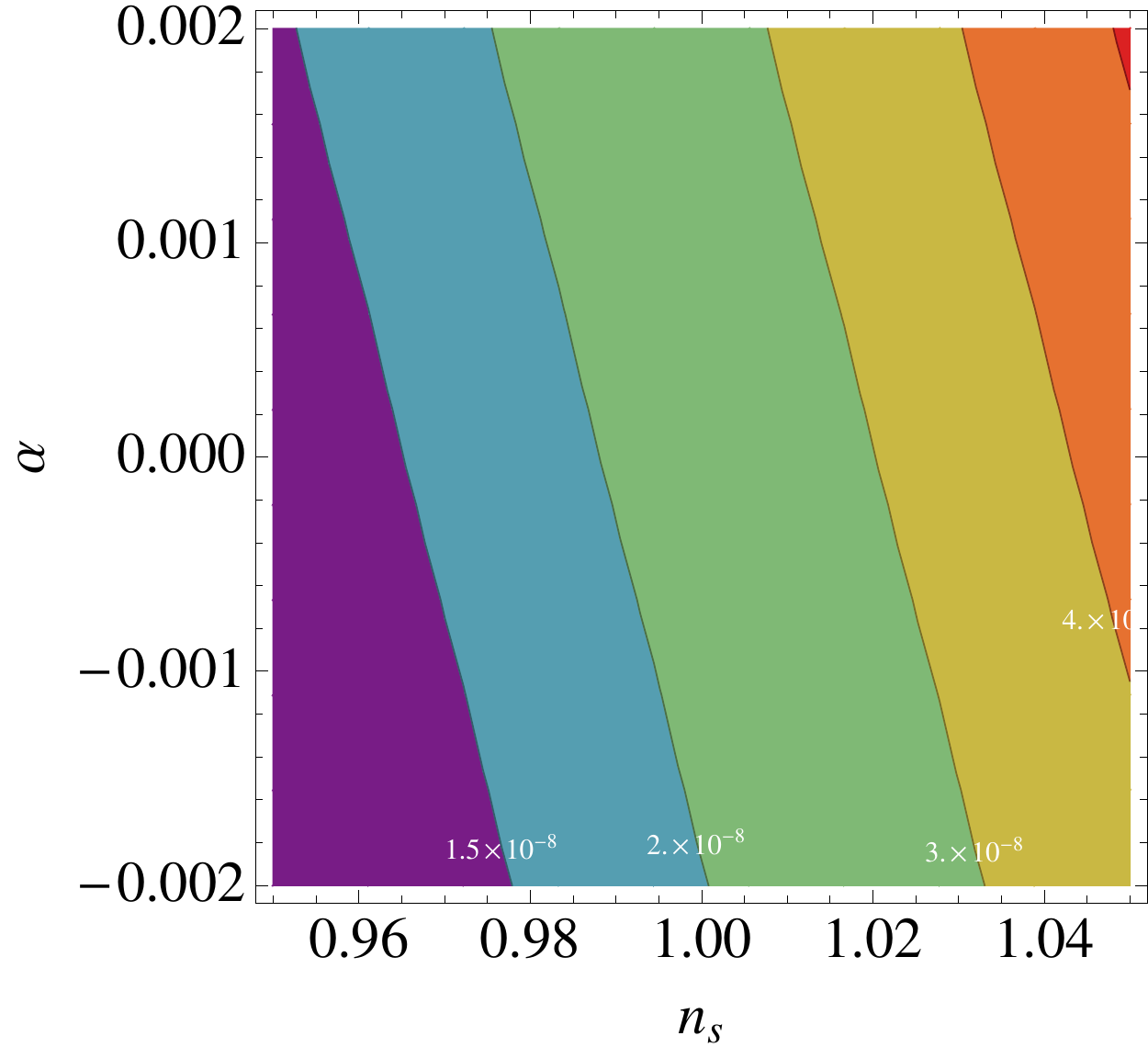}~~~~~~\includegraphics[width=8cm]{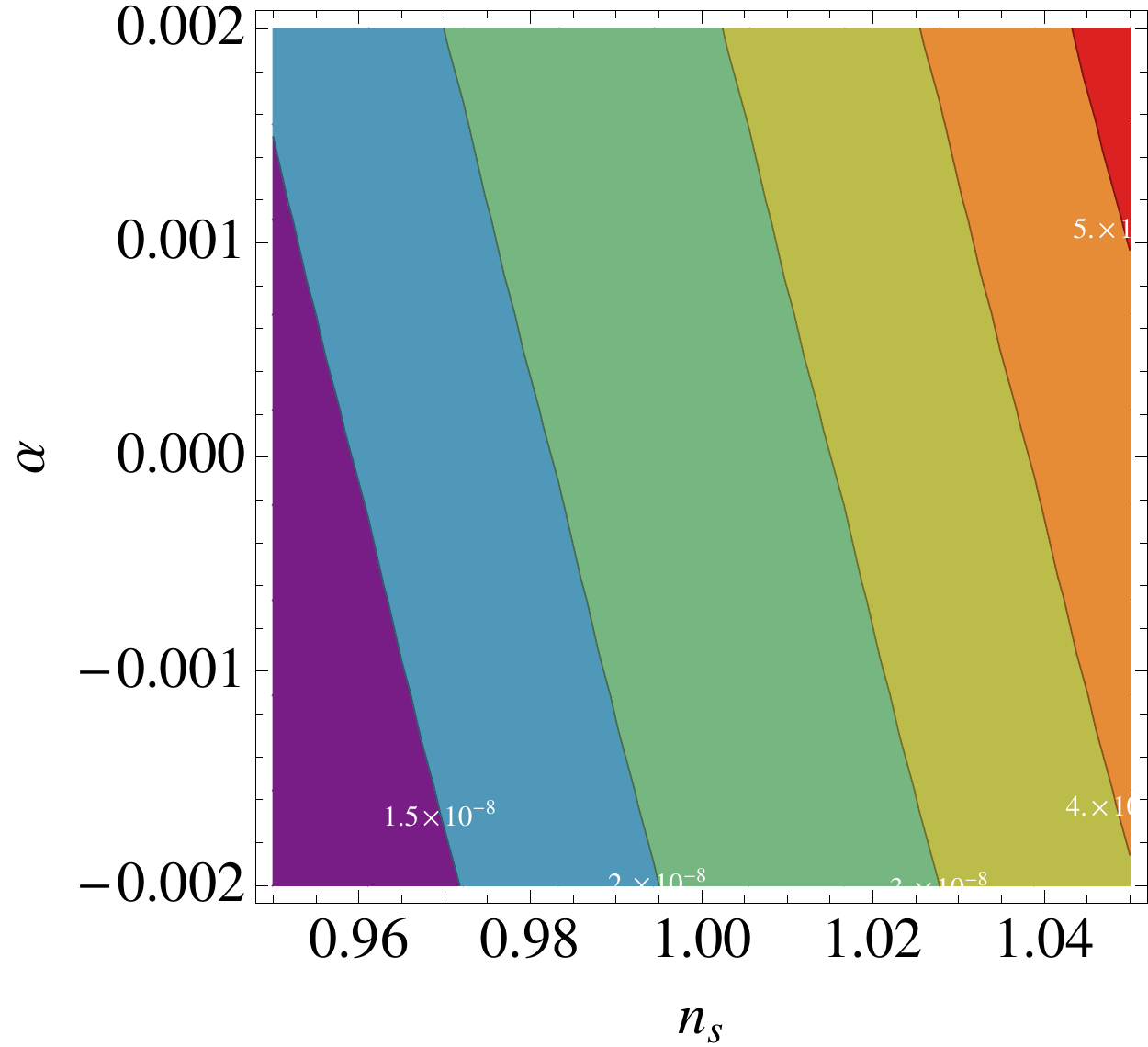}

\end{center}
\caption{The upper graph shows the predictions for the $\mu$--distortion in the case of General Relativity as a function of the spectral index $n_s$ and the running $\alpha$. The two lower graphs show the predictions for two examples of modified gravity, with the parameters taken as in Fig. 1. Since in our case the sound-speed ${\tilde c}_s$ is smaller than in the standard case, $\mu$ is predicted to be smaller for a given $n_s$ and $\alpha$.}
\end{figure}

The corresponding results for the $\mu$--distortion as a function of spectral index $n_s$ and running $\alpha$ for the evolution of ${\tilde c}_s$ are shown in Fig 2, where we also show the predictions for General Relativity which agree with {\cite{Easson}}, but with the relativistic correction of $3/4$ taken into account. As expected, the predicted $\mu$ is  smaller than in General Relativity since ${\tilde c}_s$ is smaller. For the examples shown we see that the predictions for $\mu$ are very similar, although the evolution of the sound-speed is significantly different. The sound horizon for both cases deviates less than a percent from its value in General Relativity. For the examples studied here the effects of the coupling of $\phi$ to baryons on $\mu$ are of order $10^{-9}$, similar to other, less exotic contributions. Thus the predictions for the $\mu$--distortion depend on the details of the evolution of the coupled photon-baryon fluid and therefore provide a window for modifications of gravity.

\begin{figure}
\begin{center}

\includegraphics[width=8.5cm]{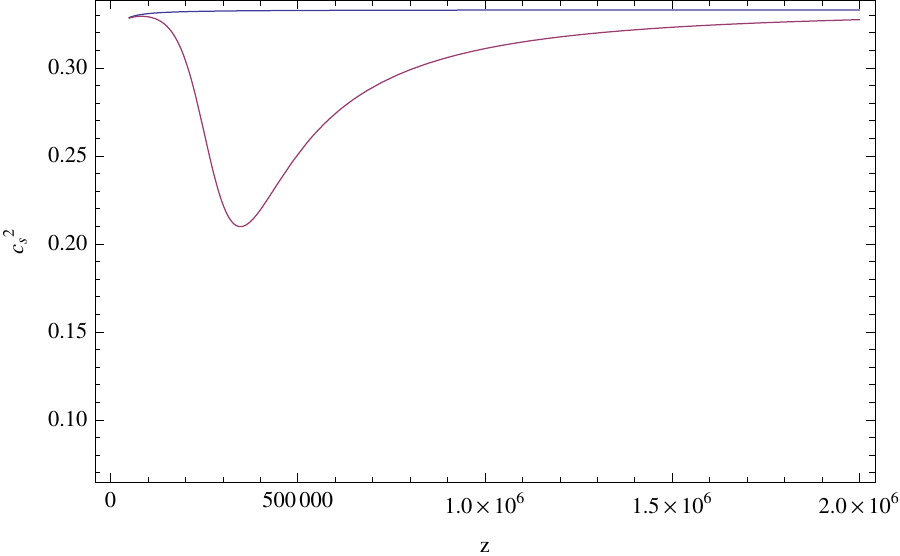}~~~~~~\includegraphics[width=8cm]{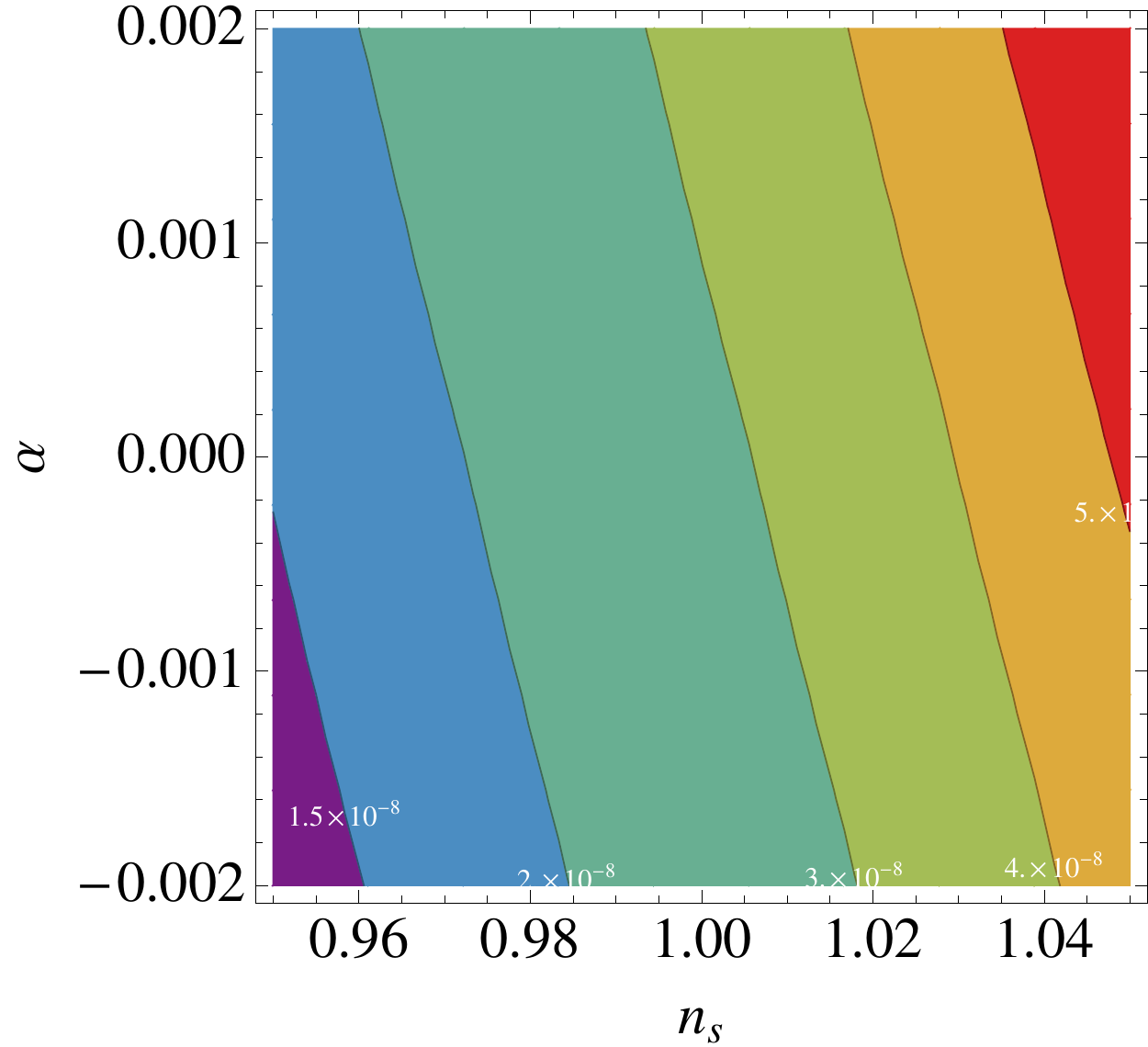}

\end{center}
\caption{In this example we have taken $b=2.5\times 10^4$, $d = 10^{-5}$, $z_0 = 2.5\times 10^5$ and $m_{\rm rat} =350$ (and $k=100$ Mpc$^{-1}$ to calculate ${\tilde c}_s$ on the left). As can be seen, the sound-speed approaches its standard value only for $z<150000$ and the deviation of $\mu$ from its value in General Relativity is very small. However, the sound horizon deviates from its value in General Relativity by 17 percent. This case is not compatible with CMB anisotropies and is shown only for illustration.}
\end{figure}

In Fig 3 we show a case for which the deviations of the effective sound-speed from $c_s$ are significant for $z>150000$, but not below. In this case the predictions for $\mu$ are very similar to in General Relativity, but the prediction for the sound horizon deviates by 17 percent. This example is ruled out by observations of the CMB anisotropies and was added just for illustration. What becomes clear from these considerations (and from eq. (\ref{devA})) is that it is the interplay between $\beta$ and $m$ (both their magnitude and evolution) which determines the predictions for $\mu$ and ${\tilde r}_s$. Even in the case of purely conformal couplings, with different choices for $\beta$ and $m$ a range of possible deviations in either ${\tilde r}_s$ or $\mu$ (or both) can be obtained. If we were to allow for a disformal coupling as well ($D\neq 0$), the results for $\mu$ and ${\tilde r}_s$ would also depend on the first derivative of the potential and, in general, on the evolution of $\phi$, at which point general statements about predictions and trends are no longer useful but instead concrete models (i.e. concrete choices for $C(\phi)$, $D(\phi)$ and $V(\phi)$) have to be studied \cite{usfuture}. However, the results above show that the $\mu$--distortion of the CMB is a useful tool to constrain modifications of gravity further. 

\section{Conclusions}
The main results of this paper can be summarised as follows: firstly, we have derived the general coupling of a fluid (relativistic or non-relativistic) to a scalar field, whose influence is described by the effective metric given in Eq. (\ref{disformalmetric}). We allowed not only for conformal but also for disformal couplings and the expressions in Section 2 are generic when the scalar field is coupled to {\it one} species only, whose equation of state is $w$. As is clear from Section 2, already in this case the evolution of the effective coupling (e.g. $Q_0/\rho_i$ in equation (\ref{Qbackgroundcoupling}) for the background evolution) can be rather complicated. The evolution of its perturbation (given in eq. (\ref{Qperturb})) is even more complicated. Therefore, in those scenarios, the coupling is in general a function of time. This opens the door to a rich phenomenology. 

Secondly, focusing on the case of the scalar field coupled to baryons, we have derived the expression for the effective sound-speed of the tightly coupled photon-baryon fluid, which differs from the expression in General Relativity. As we have pointed out, the $\mu$--distortion of the CMB can be used to constrain the evolution of ${\tilde c}_s$ and therefore constrain modifications of gravity at very high redshifts ($5\times 10^4 \leq z \leq 2\times 10^6$) and small length scales ($k>50$ Mpc$^{-1}$). Therefore, thirdly we have calculated the $\mu$--distortion for a simple case in which the coupling becomes smaller as time progresses. In this case, the $\mu$--distortion is smaller than in General Relativity, because the sound-speed is smaller. Whether this is generic requires a comprehensive analysis of different choices for the functions $C(\phi)$, $D(\phi)$ and $V(\phi)$. We will study this in future publications \cite{usfuture}. 

The $\mu$--distortion of the CMB spectrum is a useful additional probe for testing gravity and not only for the primordial power spectrum of perturbations. As we have seen in this paper, for theories with conformal couplings the coupling has to be large ($\beta > 10^3$) 
for $\mu$ to deviate significantly from its value in General Relativity. A study of disformal couplings will be presented elsewhere, in which we also study more generic theories for which the mass of the scalar field is small and modifications of the transfer functions could be important \cite{usfuture}.

\acknowledgements The work of CvdB is supported by the Lancaster-Manchester-Sheffield Consortium for Fundamental Physics under STFC grant ST/J000418/1. GS is supported by an STFC doctoral fellowship. 
We are grateful to P. Brax, J. Chluba, A.-C. Davis, D. Easson, L. Hui and J. Khoury for interesting discussions.

\section*{Appendix A: Modifications to Diffusion Damping Scale}

We follow the method outlined in \cite{Dodelson}. Damping occurs on very small scales on which the gravitational potentials are very small but the photon quadrupole is relevant. We assume that all the perturbation variables vary as $e^{i\int\omega d\eta}$ so that 
\be
\delta_b = -\frac{\theta_b}{i\omega} \qquad \text{and} \qquad \delta_{\gamma} = -\frac{4\theta_{\gamma}}{3i\omega}.
\ee
Equation (\ref{sHthetabdot}) then becomes
\be
i\omega\theta_b = \frac{\dot{\tau}}{R}(\theta_{\gamma}-\theta_{b})-\frac{k^{2}{\cal F}}{i\omega}\theta_{b}.
\ee
Rearranging, we find to second order in $\dot{\tau}^{-1}$
\begin{align}
\theta_b = \, & \left[1+\frac{i\omega R}{\dot{\tau}}\left(1-\frac{k^{2}{\cal F}}{\omega^{2}}\right)\right]^{-1}\theta_{\gamma} \nonumber \\ \approx \, & \left[1-\frac{i\omega R}{\dot{\tau}}\left(1-\frac{k^{2}{\cal F}}{\omega^{2}}\right)-\left(\frac{\omega R}{\dot{\tau}}\right)^{2}\left(1-\frac{k^{2}{\cal F}}{\omega^{2}}\right)^{2}\right]\theta_{\gamma}~.
\end{align}
Inserting this in equation (\ref{sHthetagammadot}) gives
\be
i\omega = -\frac{k^{2}}{3i\omega}-\frac{16k^{2}}{45\dot{\tau}}+\dot{\tau}\left(\left[1-\frac{i\omega R}{\dot{\tau}}\left(1-\frac{k^{2}{\cal F}}{\omega^{2}}\right)-\left(\frac{\omega R}{\dot{\tau}}\right)^{2}\left(1-\frac{k^{2}{\cal F}}{\omega^{2}}\right)^{2}\right]-1\right)
\ee
and after collecting terms we obtain
\be
\omega^{2} = k^{2}c_{s}^{2}(1+3R{\cal F})+\frac{i\omega k^{2}c_{s}^{2}}{\dot{\tau}}\left[\frac{16}{15}+\frac{3\omega^{2}R^{2}}{k^{2}}-6R^{2}{\cal F}+\frac{3k^{2}R^{2}}{\omega^{2}}{\cal F}^{2}\right].
\ee
We can recognise the first term on the right-hand side as $k^{2}{\tilde c}_{s}^{2}$, and this allows us to write
\begin{align}
\omega = \, & k{\tilde c}_{s}+\frac{i\omega kc_{s}^{2}}{2\dot{\tau}{\tilde c}_{s}}\left[\frac{16}{15}+\frac{3\omega^{2}R^{2}}{k^{2}}-6R^{2}{\cal F}+\frac{3k^{2}R^{2}}{\omega^{2}}{\cal F}^{2}\right] \nonumber \\ = \, & k{\tilde c}_{s}+\frac{i k^{2}c_{s}^{2}}{2\dot{\tau}}\left[\frac{16}{15}+\frac{R^{2}}{1+R}\left(1-3(2+R){\cal F}+\frac{3(1+R)}{{\tilde c}_{s}^{2}}{\cal F}^{2}\right)\right],
\end{align}
where, in the second equality, we have inserted the zero order part of the $1/\dot{\tau}$ expansion ($\omega_0 = k{\tilde c}_{s}$) into the first order correction. Now using this expression in our ansatz for $\delta_\gamma$ we find
\be
\delta_\gamma \propto e^{ik{\tilde r}_{s}}e^{-k^{2}/{\tilde k}_{D}^{2}}
\ee
with
\be
\frac{1}{{\tilde k}_D^2} = \int_z^{\infty}\frac{dz(1+z)}{6H(1+R)n_{e}\sigma_T}\left[\frac{16}{15}+\frac{R^{2}}{1+R}\left(1-3(2+R){\cal F}+\frac{3(1+R)}{{\tilde c}_{s}^{2}}{\cal F}^{2}\right)\right].
\ee

\end{document}